\documentclass[aps,pre,reprint,showpacs,superscriptaddress,10pt]{revtex4-1}
\usepackage{amsmath,amssymb,amsfonts,graphicx,bm,dcolumn,mathrsfs}
\usepackage{color}
\usepackage{ulem}

\begin{document}

\newcommand{\dif}{d}
\newcommand{\ee}{e}
\newcommand{\ii}{i}
\newcommand{\kB}{k_B}
\newcommand{\sub}[1]{\text{#1}}
\newcommand{\vect}[1]{\bm{#1}}

\title{Long-range interacting systems in the unconstrained ensemble}

\author{Ivan Latella}
\email{ivan.latella@institutoptique.fr}
\affiliation{Laboratoire Charles Fabry, UMR 8501, Institut d'Optique, CNRS, Universit\'e Paris-Saclay, 2 Avenue Augustin Fresnel,
91127 Palaiseau Cedex, France}

\author{Agust\'in P\'erez-Madrid}
\email{agustiperezmadrid@ub.edu}
\affiliation{Departament de F\'{i}sica de la Mat\`{e}ria Condensada, Facultat de F\'{i}sica, Universitat de Barcelona,
Mart\'{i} i Franqu\`{e}s 1, 08028 Barcelona, Spain}

\author{Alessandro Campa}
\email{alessandro.campa@iss.infn.it}
\affiliation{Complex Systems and Theoretical Physics Unit, Health and Technology Department, 
Istituto Superiore di Sanit\`{a}, and INFN Roma 1, Viale Regina Elena 299, 00161 Roma, Italy}

\author{Lapo Casetti}
\email{lapo.casetti@unifi.it}
\affiliation{Dipartimento di Fisica e Astronomia and CSDC, Universit\`a di Firenze,\\and INFN, Sezione di Firenze,
via G.\ Sansone 1, 50019 Sesto Fiorentino (FI), Italy}
\affiliation{INAF-Osservatorio Astrofisico di Arcetri, Largo E. Fermi 5, 50125 Firenze, Italy}

\author{Stefano Ruffo}
\email{ruffo@sissa.it}
\affiliation{SISSA, via Bonomea 265, ISC-CNR and INFN, 34136 Trieste, Italy}

\begin{abstract}
Completely open systems can exchange heat, work, and matter with the environment. While energy, volume, and number of particles fluctuate under completely open conditions, the equilibrium states of the system, if they exist, can be specified using the temperature, pressure, and chemical potential as control parameters. 
The unconstrained ensemble is the statistical ensemble describing completely open systems and the replica energy is the appropriate free energy for these control parameters from which the thermodynamics must be derived.
It turns out that macroscopic systems with short-range interactions cannot attain equilibrium configurations in the unconstrained ensemble, since temperature, pressure, and chemical potential cannot be taken as a set of independent variables in this case.
In contrast, we show that systems with long-range interactions can reach states of thermodynamic equilibrium in the unconstrained ensemble. 
To illustrate this fact, we consider a modification of the Thirring model and compare the unconstrained ensemble with the canonical and grand canonical ones: the more the ensemble is constrained by fixing the volume or number of particles, the larger the space of parameters defining the equilibrium configurations.
\end{abstract}


\maketitle

\section{\label{intro}Introduction}

Systems with long-range interactions may display a notable thermodynamic behavior that distinguish them from those where the interactions are
short-ranged~\cite{Campa_2014,Bouchet_2010,Levin_2014}. Long-range interactions are those for which the range is comparable with
the size of the system, regardless of how large the system is, as it occurs, for instance, in self-gravitating systems~\cite{Antonov_1962,Lynden-Bell_1968,
Thirring_1970,Padmanabhan_1990,Lynden-Bell_1999,Chavanis_2002,deVega_2002,Chavanis_2006}, plasmas~\cite{Kiessling_2003,Nicholson_1992},
fluid dynamics~\cite{Chavanis_2002_b,Robert_1991}, and some spin systems~\cite{Mori_2013}. 
Because of these interactions, the system may remain trapped in nonequilibrium quasi-stationary states~\cite{Benetti_2012} whose lifetimes depend on the number of particles and diverge in the limit $N\to\infty$. For large but finite number of particles in a very long time limit, however, the system eventually evolves towards states of thermodynamic equilibrium~\cite{Yamaguchi_2004} that can be described within the usual framework of ensemble theory. At equilibrium, such systems may present ensemble inequivalence~\cite{Thirring_1970,Ellis_2000,Barre_2001,Bouchet_2005} associated to anomalies in the concavity of the thermodynamic potentials~\cite{Chomaz_2002,Bouchet_2005}.
In particular, a negative heat capacity in the microcanonical ensemble~\cite{Lynden-Bell_1968,Thirring_1970,Schmidt_2001} is due to an
anomaly in the concavity of the entropy as a function of the energy.
In addition, systems with long-range interactions, in general, do not obey the usual Gibbs-Duhem equation~\cite{Latella_2013,Latella_2015} (see Ref.~\cite{Velazquez_2016} for a discussion regarding self-gravitating systems).
This fact is deeply related with the nonadditive character that long-range interactions confer to these systems~\cite{Latella_2015}.
Actually, nonadditivity is responsible for the remarkable behavior observed in long-range interacting systems~\cite{Campa_2014}, as well as for the physics behind the subject we discuss in this paper: the appearance of a new equilibrium statistical ensemble in the thermodynamic limit.

In this paper we focus on the thermodynamics of long-range interacting systems under completely open conditions.
A completely open system can exchange heat, work, and matter with its surroundings. That is, the energy, volume, and number of particles of the system fluctuate under completely open conditions. Thus, the control parameters that specify the thermodynamic state of the system are temperature, pressure, and chemical potential. These control parameters are properties of a suitable reservoir that weakly interacts with the system, in the sense that, by means of some mechanism, it supplies heat, work, and matter to the system, but can be assumed not to be coupled by long-range interactions. 
It is worth mentioning that under these conditions the system is not enclosed by rigid walls. Instead, the system could be confined by an external field exerting pressure on it; e.g., tidal forces produced by surrounding objects in self-gravitating systems or magnetic fields acting on plasmas~\cite{Levin_2008}.
Furthermore, the unconstrained ensemble is the statistical ensemble that describes a completely open system (in the literature, this ensemble is also termed generalized~\cite{Hill_1987}). It was introduced by Guggenheim~\cite{Guggenheim_1939}, but it did not receive much attention due to its lack of application in standard macroscopic systems. If the interactions in the system are short-ranged, the free energy associated to this ensemble is vanishingly small when the number of particles is large~\cite{Guggenheim_1939}. This is a consequence of the fact that temperature, pressure, and chemical potential cannot be treated as independent variables 
because they are related by the Gibbs-Duhem equation. Thus they are not, taken together, suitable control parameters for macroscopic systems with short-range interactions.

In case that the limit of large number of particles is not assumed, however, the situation changes, as pointed out by Hill~\cite{Hill_1963}. Small
systems may have an extra degree of freedom which permits that equilibrium configurations in completely open conditions can be
realized~\cite{Hill_1963,Hill_1998,Hill_2001,Hill_2002}. Nevertheless, when the size of the system is increased, the usual behavior of macroscopic
systems is obtained~\cite{Hill_1963,Schnell_2011,Schnell_2012}, as long as the interactions remain short-ranged.

Our aim here is to show that equilibrium configurations with a large number of particles may be realized in the unconstrained ensemble if the interactions in the system are long-ranged. Moreover, we argue that the replica energy~\cite{Latella_2015} is the appropriate free energy defining such configurations in this ensemble. Thus, in Sec.~\ref{unconstrained_ensemble}, we will first describe this ensemble and make the appropriate
connection with the thermodynamics. In Sec.~\ref{modified_Thirring_model} we introduce a solvable model, whose thermodynamics in the unconstrained
ensemble is discussed in Sec.~\ref{completely_open_model}. In Secs.~\ref{grandcanonical_model} and \ref{canonical_model}, this model is studied in
the grand canonical and canonical ensembles, respectively. In addition, some thermodynamic relations associated with the replica energy are
analyzed in Sec.~\ref{thermodynamic_relations}, and our conclusions are presented in Sec.~\ref{conclusions}.

\section{\label{unconstrained_ensemble}The unconstrained ensemble}

To make our point clear, it is worth to briefly review the thermostatistics of completely open systems by establishing its connection to the
unconstrained ensemble. The description of the unconstrained ensemble presented in this section is based on Refs.~\cite{Hill_1987} and \cite{Hill_1963}, which the reader is referred to for an extended discussion (see also~\cite{Guggenheim_1939,Guggenheim_1967,Hill_2001,Planes_2002}).

The energy, volume, and number of particles of a completely open system are quantities that fluctuate due the interaction of the system with its surroundings. Let us consider the probability $p_i(V,N)$ of the configuration of a system that is found in the state $i$ with energy $E_i$, and
possesses $N$ particles in a volume $V$. This probability is an exponential of the form~\cite{Hill_1987}
\begin{equation}
p_i(V,N)=\frac{\exp\left[-\alpha N-\beta E_i(V,N)-\gamma V\right]}{\Upsilon},
\label{probability}
\end{equation}
where $\alpha$, $\beta$, and $\gamma$ are parameters that will be identified below and $\Upsilon$ is the associated partition function given by
\begin{equation}
\Upsilon=\sum_{i,V,N} \exp\left[-\alpha N-\beta E_i(V,N)-\gamma V\right].
\end{equation}
Here discrete variables are used for simplicity. The ensemble average $\bar{E}$ of the internal energy $E$ is thus obtained as
$\bar{E}=\sum_{i,V,N}E_i(V,N)p_i(V,N)$, while the average number of particles $\bar{N}$ and the average volume $\bar{V}$ read
$\bar{N}=\sum_{i,V,N}Np_i(V,N)$ and $\bar{V}=\sum_{i,V,N}Vp_i(V,N)$, respectively. Now let us consider an infinitesimal change of the average
internal energy in terms of changes in the probability,
\begin{equation}
\dif\bar{E}=\sum_{i,V,N} E_i(V,N)\dif p_i(V,N),
\label{dif_E}
\end{equation}
where the coefficients $E_i(V,N)$ are assumed constant.
Hence, using Eq.~(\ref{probability}) to express $E_i(V,N)$ and the condition $\sum_{i,V,N}\dif p_i(V,N)=0$, Eq.~(\ref{dif_E}) can be rewritten as
\begin{equation}
\dif \bar{E}=-\frac{1}{\beta}\dif\left[\sum_{i,V,N}p_i(V,N)\ln p_i(V,N) \right]-\frac{\alpha}{\beta}\dif\bar{N}-\frac{\gamma}{\beta}\dif\bar{V}.
\label{dif_E_2}
\end{equation}
If Eq.~(\ref{dif_E_2}) is compared with the thermodynamic equation
\begin{equation}
\dif \bar{E}=T\dif S-P\dif\bar{V}+\mu\dif\bar{N},
\label{Gibbs_equation}
\end{equation}
where $T$ is the temperature, $P$ is the pressure, and $\mu$ is the chemical potential,
one then recognizes $\kB T=1/\beta$, $\mu=-\alpha/\beta$, $P=\gamma/\beta$, and the entropy 
\begin{equation}
S=-\kB\sum_{i,V,N}p_i(V,N)\ln p_i(V,N),
\label{entropy_completely_open}
\end{equation}
where $\kB$ is the Boltzmann constant.
Therefore, the probability (\ref{probability}) leads to the correct thermostatistical description of completely open systems. Furthermore,
substituting Eq.~(\ref{probability}) in Eq.~(\ref{entropy_completely_open}) with the previous identifications, one gets
\begin{equation}
\mathscr{E}=\bar{E}-TS+P \bar{V}-\mu \bar{N},
\label{replica_energy_completely_open}
\end{equation}
where we have introduced the free energy
\begin{equation}
\mathscr{E}(T,P,\mu)=-\kB T\ln\Upsilon(T,P,\mu), 
\end{equation}
called replica energy~\cite{Latella_2015} or subdivision potential~\cite{Hill_1963}.
By differentiation of Eq.~(\ref{replica_energy_completely_open}) and using Eq.~(\ref{Gibbs_equation}), one obtains
\begin{equation}
\dif\mathscr{E}=-S\dif T+ \bar{V}\dif P-\bar{N}\dif\mu.
\label{generalized_Gibbs_Duhem}
\end{equation}
Notice that, as a consequence of Eq.~(\ref{generalized_Gibbs_Duhem}), one has
\begin{eqnarray}
\left(\frac{\partial\mathscr{E}}{\partial T}\right)_{P,\mu}&=&-S,\label{entropy_relation}\\ 
\left(\frac{\partial\mathscr{E}}{\partial P}\right)_{T,\mu}&=&\bar{V},\label{volume_relation}\\
\left(\frac{\partial\mathscr{E}}{\partial \mu}\right)_{T,P}&=&-\bar{N}.\label{N_relation}
\end{eqnarray}
Moreover, the unconstrained partition function can be written as
\begin{equation}
\Upsilon(T,P,\mu)=\sum_{V,N}Z(T,V,N)\ e^{\mu N/(\kB T)}e^{-PV/(\kB T)},
\label{completely_open_partition_function}
\end{equation}
where
\begin{equation}
Z(T,V,N)=\sum_ie^{-E_i(V,N)/(\kB T)} 
\end{equation}
is the canonical partition function. From Eq.~(\ref{completely_open_partition_function}), it is straightforward to connect $\Upsilon$ with the partition function of another ensemble, e.g., one has
\begin{equation}
\Upsilon(T,P,\mu)=\sum_{V}\Xi(T,V,\mu)\ e^{-PV/(\kB T)},
\end{equation}
where $\Xi(T,V,\mu)$ is the grand canonical partition function.

For macroscopic systems with short-range interactions in the thermodynamic limit, the internal energy and the other thermodynamic potentials
are linear homogeneous functions of the
extensive variables and thus, from Eq.~(\ref{replica_energy_completely_open}), one obtains $\mathscr{E}=0$, since in this case the Gibbs free
energy $G=\bar{E}-TS+P \bar{V}$ is equal to $\mu \bar{N}$. We note that in this case it does
not mean that $\Upsilon=1$ but that $\mathscr{E}$ is negligible in the thermodynamic limit~\cite{Hill_1987}. Thus, for this kind of systems,
Eq.~(\ref{generalized_Gibbs_Duhem}) reduces to the well-known Gibbs-Duhem equation. The interesting fact is that
equation (\ref{replica_energy_completely_open}) indicates the route to deal with the more general situation in which the internal energy is
not a linear homogeneous function of $S$, $V$, and $N$, as it happens with systems with long-range interactions. 
This is the case in which here
we are concerned with, for which the replica energy is, in general, different from zero.

The above considerations are directly related to the fact that for an ensemble of $\mathscr{N}$ replicas of the system with total energy $E_\text{t}=\mathscr{N}E$, entropy $S_\text{t}=\mathscr{N}S$, volume $V_\text{t}=\mathscr{N}V$, and number of particles $N_{\text{t}}=\mathscr{N}N$, $\mathscr{E}$ can be interpreted as the energy gained by the ensemble when the number of replicas changes holding $S_\text{t}$, $V_\text{t}$ and $N_{\text{t}}$ constant.
Formally~\cite{Hill_1963},
\begin{equation}
\mathscr{E}=\left(\frac{\partial E_\text{t}}{\partial\mathscr{N}}\right)_{S_\text{t}, V_\text{t},N_{\text{t}}}. 
\label{eq_E}
\end{equation}
When the single-system energy $E$ is a linear homogeneous function of $S$, $V$, and $N$ (additive systems), one has $E_\text{t}=\mathscr{N}E(S_\text{t}/\mathscr{N},V_\text{t}/\mathscr{N},N_\text{t}/\mathscr{N})=E(S_\text{t},V_\text{t},N_\text{t})$ and therefore, Eq.~(\ref{eq_E}) leads to $\mathscr{E}=0$. If $E$ is not a linear homogeneous function of $S$, $V$, and $N$ (nonadditive systems), Eq.~(\ref{eq_E}) then requires $\mathscr{E}\neq0$~\cite{Latella_2015}.

Finally, we want to stress that Eqs.~(\ref{replica_energy_completely_open}) and (\ref{generalized_Gibbs_Duhem})-(\ref{N_relation}) are relations at a
thermodynamic level. 
That is, for any other ensemble, corresponding to the actual physical conditions
specified by the given control parameters, an analogous set of equations can be obtained
from the appropriate free energy~\cite{Hill_1963}. The replica energy will emerge also in those
cases, although it will not be the free energy associated to the partition function
for that ensemble. However, since in any case $\mathscr{E}$ vanishes when the system is additive, it can always be seen as a measure of the nonadditivity of the system~\cite{Latella_2015}.

\section{\label{modified_Thirring_model}The modified Thirring model}

We introduce here a model that, as discussed in the following, can attain equilibrium configurations under completely open conditions.
Consider a system of $N$ particles of mass $m$ enclosed in a volume $V$ with a Hamiltonian of the form
\begin{equation}
\mathcal{H}(\vect{p},\vect{q})=\sum_{i=1}^N\frac{|\vect{p}_i|^2}{2m}+\sum_{i> j}^N\phi(\vect{q}_i,\vect{q}_j),
\label{Hamiltonian}
\end{equation}
where $\vect{p}_i\in\mathbb{R}^3$ and $\vect{q}_i\in\mathbb{R}^3$ are the momentum and position of particle $i$, respectively,
$\phi(\vect{q}_i,\vect{q}_j)$ is the interaction potential, $\vect{p}=(\vect{p}_1,\ldots,\vect{p}_N)$, and $\vect{q}=(\vect{q}_1,\ldots,\vect{q}_N)$.
The interactions in this model are defined by
\begin{equation}
\phi(\vect{q}_i,\vect{q}_j)=-2\nu\left[ \theta_{V_0}(\vect{q}_i) \theta_{V_0}(\vect{q}_j)+b\theta_{V_1}(\vect{q}_i) \theta_{V_1}(\vect{q}_j)\right],
\label{interaction_potential}
\end{equation}
where $\nu>0$ and $b$ are constants, and we have introduced the functions
\begin{equation}
\begin{split}
&\theta_{V_0}(\vect{q}_i)=
\begin{cases}
1 & \text{if } \vect{q}_i\in V_0\\
0 & \text{if } \vect{q}_i\notin V_0
\end{cases},\\
&\theta_{V_1}(\vect{q}_i)=
\begin{cases}
1 & \text{if } \vect{q}_i\notin V_0\\
0 & \text{if } \vect{q}_i\in V_0
\end{cases}. 
\end{split}
\end{equation}
Here $V_0$ is the volume of an internal region of the system such that $V_0< V$, and $V_1$ is the volume of the region outside $V_0$ so that
$V_1= V-V_0$. In this way, $V_0$ is a parameter of the interaction potential that does not depend on the state of the system. Hence, the total
potential energy in the large $N$ limit is given by
\begin{equation}
\hat{W}(N_0,N_1)\equiv \sum_{i> j}^N\phi(\vect{q}_i,\vect{q}_j)=-\nu\left( N_0^2+bN_1^2\right),
\end{equation}
where $N_0$ is the number of particles in $V_0$ and $N_1=N-N_0$ is the number of particles in $V_1$ for a given configuration.
Notice that $N_0=N_0(\vect{q})$ and $N_1=N_1(\vect{q})$, so that $\hat{W}$ depends implicitly on the position of all particles.
This model is a modification of the Thirring model~\cite{Thirring_1970}, which is obtained in the particular case $b=0$.
Furthermore, the model is solvable in the different statistical ensembles, and below we focus on the unconstrained ensemble and compare it
with the canonical and grand canonical cases. In the following it will become clear why we consider this modification of the model.

\subsection{\label{completely_open_model}The unconstrained ensemble}

Here we study the modified Thirring model in the unconstrained ensemble. Thus, we will assume that the system is in contact
with a reservoir characterized by fixed temperature $T$, pressure $P$, and chemical potential $\mu$.

\begin{figure}
\includegraphics[scale=1]{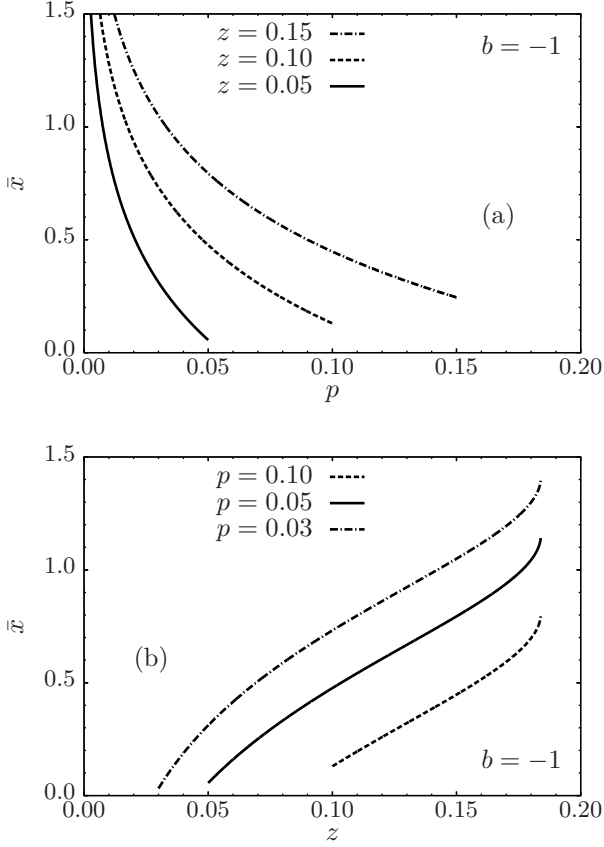}
\caption{Reduced number of particles $\bar{x}$ for the modified Thirring model with $b=-1$ in the unconstrained ensemble:
in (a), as a function of the reduced pressure $p$ with constant fugacity $z$, and, in (b), as a function of the fugacity with constant reduced pressure.
We remind that in this and in the following four figures we have, as requested by the equilibrium condition in the unconstrained ensemble, $p<z$ and
$z<1/(2e)$.}
\label{graph_p_x_z}
\end{figure}

Consider the canonical partition function for this model, which is given by
\begin{eqnarray}
Z(T,V,N)&=&\int\frac{\dif^{3N}\vect{q}\ \dif^{3N}\vect{p}}{h^{3N}N!}\ e^{-\beta\mathcal{H}(\vect{p},\vect{q})}\nonumber\\
&=& \int\frac{\dif^{3N}\vect{q}}{N!}\ \frac{e^{-\beta \hat{W}(N_0,N_1)}}{\lambda_T^{3N}},
\label{partition_fucntion}
\end{eqnarray}
where $\lambda_T=h/\sqrt{2\pi m T}$ is the thermal wave length, $h$ being a constant, and hereafter we take units where $\kB=1$. Thus, using
Eq.~(\ref{completely_open_partition_function}), the unconstrained partition function becomes (we take $V$ as a continuous variable)
\begin{eqnarray}
\Upsilon(T,P,\mu)&=&\int \dif V\sum_N\int\frac{\dif^{3N}\vect{q}}{N!}\nonumber\\*
&&\times\lambda_T^{-3N}\ e^{-\beta\hat{W}(N_0,N_1)} e^{\beta\mu N}e^{-\beta PV}.
\label{completely_open_partition_function_2} 
\end{eqnarray}
Following Thirring's method~\cite{Thirring_1970}, the partition function can be computed by replacing the integral
over coordinates with a sum over the occupation numbers in each region of the system, 
\begin{equation}
\int\frac{\dif^{3N}\vect{q}}{N!} \rightarrow\sum_{N_0,N_1}\delta_{N,N_0+N_1}\frac{V_0^{N_0}}{N_0!}\frac{V_1^{N_1}}{N_1!} ,
\label{Thirring_method}
\end{equation}
where the sum runs over all possible values of $N_0$ and $N_1$ and the Kronecker $\delta$ enforces the condition $N=N_0+N_1$. This leads to
\begin{equation}
\Upsilon(T,P,\mu)=\int \dif V\sum_{N_0,N_1}\ e^{-\beta\hat{\mathscr{E}}(T,P,\mu,V,N_0,N_1)},
\label{completely_open_partition_function_3}
\end{equation}
where we have introduced (from now on we omit writing down explicitly the dependence on the control parameters)
\begin{eqnarray}
\hat{\mathscr{E}}(V,N_0,N_1)&=&\hat{W}(N_0,N_1)+PV-T\sum_kN_k\nonumber\\*
&&+T\sum_kN_k\left[\ln\left(N_k\frac{\lambda_T^3}{V_k}\right)-\frac{\mu}{T}\right],
\label{replica_energy_hat}
\end{eqnarray}
with $k=0,1$, and we have used Stirling's approximation.
Since the replica energy is given by $\mathscr{E}=- T\ln\Upsilon$, using the saddle-point approximation one obtains
\begin{equation}
\mathscr{E}=\inf_{\{V,N_0,N_1\}} \hat{\mathscr{E}}(V,N_0,N_1).
\label{replica_energy_minimization}
\end{equation}
Thus, minimization with respect to $V$, $N_0$, and $N_1$ requires that
\begin{eqnarray}
P&=&\frac{T\bar{N}_1}{\bar{V}-V_0},\label{P_open}\\
\mu&=&-2\nu \bar{N}_0+T\ln\left(\frac{\bar{N}_0}{V_0}\lambda_T^3\right),\label{mu_1_open}\\
\mu&=&-2b\nu \bar{N}_1+T\ln\left(\frac{\bar{N}_1}{\bar{V}-V_0}\lambda_T^3\right),\label{mu_2_open}
\end{eqnarray}
where $\bar{V}$, $\bar{N}_0$, and $\bar{N}_1$ are the values of the volume and the number of particles in each region that minimizes the replica
energy. It is clear that $\bar{V}$, $\bar{N}_0$, and $\bar{N}_1$ are functions of $T$, $P$, and $\mu$, whose dependence is implicit through
Eqs.~(\ref{P_open})-(\ref{mu_2_open}). The mean value of the total number of particles is then given by $\bar{N}=\sum_k\bar{N}_k$. Moreover, the
total potential energy for equilibrium configurations in the unconstrained ensemble becomes
\begin{equation}
W\equiv\hat{W}(\bar{N}_0,\bar{N}_1)=-\nu\left( \bar{N}_0^2+b\bar{N}_1^2\right).
\end{equation}
Hence, notice that from Eqs.~(\ref{mu_1_open}) and (\ref{mu_2_open}) one obtains
\begin{equation}
\mu \bar{N}=T\sum_k\bar{N}_k\ln\left(\frac{\bar{N}_k}{V_k}\lambda_T^3\right)+ 2 W ,
\end{equation}
with $V_1=\bar{V}-V_0$. Therefore, using Eqs.~(\ref{replica_energy_hat}) and (\ref{replica_energy_minimization}), the replica energy takes the form
\begin{equation}
\mathscr{E}=-W+P^{(\text{e})}\bar{V},
\label{replica_energy_2}
\end{equation}
as given in~\cite{Latella_2015}, where $P^{(\text{e})}=P-\bar{N}T/\bar{V}$ is the excess pressure.

\begin{figure}
\includegraphics[scale=1]{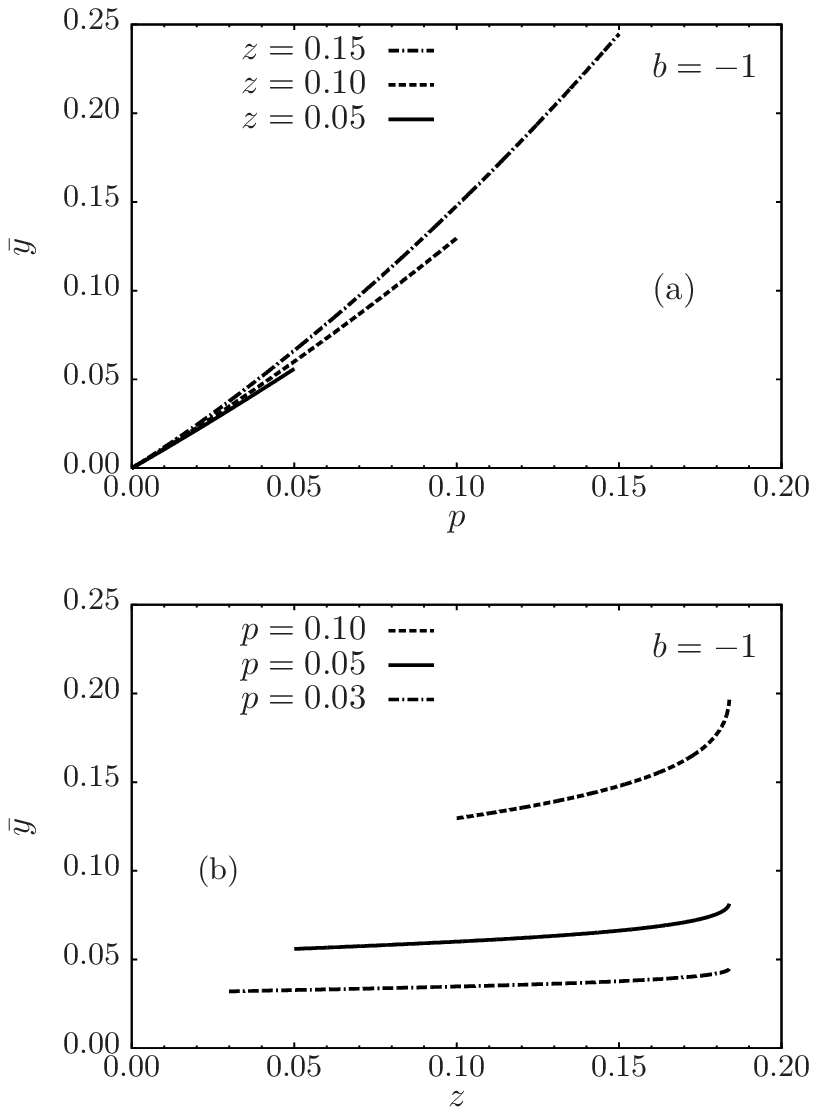}
\caption{Reduced density $\bar{y}$ for the modified Thirring model with $b=-1$ in the unconstrained ensemble:
in (a), as a function of the reduced pressure $p$ with constant fugacity $z$, and, in (b), as a function of the fugacity with constant reduced pressure.}
\label{graph_p_y_z}
\end{figure}

We introduce the dimensionless variables
\begin{equation}
v=\frac{V-V_0}{V_0},\qquad x_0=\frac{\nu N_0}{T},\qquad x_1=\frac{\nu N_1}{T},
\label{v_x_0_x_1}
\end{equation}
which will be denoted as $\bar{v}$, $\bar{x}_0$, and $\bar{x}_1$ when evaluated at $\bar{V}$, $\bar{N}_0$, and $\bar{N}_1$, respectively.
In addition, we define the reduced pressure $p$ and the relative fugacity $z$ by
\begin{equation}
p\equiv\frac{\nu V_0}{T^2}P,\qquad z\equiv e^{(\mu-\mu_0)/T},
\label{p_z}
\end{equation}
where 
\begin{equation}
\mu_0=T\ln\left(\frac{T\lambda_T^3}{\nu V_0}\right). 
\end{equation}
Since controlling $T$, $P$, and $\mu$ is equivalent to controlling $T$, $p$, and $z$, the latter set of variables can be taken as the set of
control parameters in this ensemble. Using these variables, from Eqs.~(\ref{P_open})-(\ref{mu_2_open}) one obtains
\begin{eqnarray}
\bar{x}_0&=&ze^{2\bar{x}_0},\label{x_0_open}\\
\bar{x}_1(p,z)&=&\frac{1}{2b}\ln \left(\frac{p}{z}\right),\label{x_1_open}\\
\bar{v}(p,z)&=&\frac{1}{2bp}\ln \left(\frac{p}{z}\right)\label{v},
\end{eqnarray}
where Eq.~(\ref{x_0_open}) defines implicitly  $\bar{x}_0=\bar{x}_0(z)$.
Furthermore, let us consider the reduced replica energy $\varphi_U=\nu\mathscr{E}/T^2$ and the function $\hat{\varphi}_U=\nu\hat{\mathscr{E}}/T^2$,
where, using Eqs.~(\ref{v_x_0_x_1}) and (\ref{p_z}), the latter can be written as
\begin{eqnarray}
\hat{\varphi}_U(v,x_0,x_1)&=& x_0\left[\ln\left(\frac{x_0}{z}\right)-1\right] -x_0^2 +p(v+1)\nonumber\\*
&&+ x_1\left[\ln\left(\frac{x_1}{vz}\right)-1\right]-bx_1^2. 
\end{eqnarray}
We note that with the dimensionless quantities, i.e., in the reduced replica energy and in Eqs.~(\ref{x_0_open})-(\ref{v}),
the temperature does not appear; therefore $T$ acts as a simple scaling factor.
The condition (\ref{replica_energy_minimization}) becomes
\begin{equation}
\varphi_U(\bar{v},\bar{x}_0,\bar{x}_1)=\inf_{\{v,x_0,x_1\}} \hat{\varphi}_U(v,x_0,x_1). 
\end{equation}
Since the Hessian matrix $H_U$ associated to $\hat{\varphi}_U$ at the stationary point $(\bar{v},\bar{x}_0,\bar{x}_1)$ takes the form
\begin{equation}
H_U=
\begin{pmatrix}
2b p^2\ln^{-1}(p/z) & 0 & -2b p\ln^{-1}(p/z)\\
0 & 1/\bar{x}_0-2  & 0 \\
-2b p\ln^{-1}(p/z) & 0 & 2b \left[\ln^{-1}(p/z)-1\right]
\end{pmatrix},
\end{equation}
one infers that $\bar{v}$, $\bar{x}_0$, and $\bar{x}_1$ lead to a minimum of replica energy when
\begin{equation}
\bar{x}_0<1/2,\qquad b<0,\qquad p<z. 
\end{equation}
The last two inequalities guarantee, from Eqs. (\ref{x_1_open}) and (\ref{v}), that $\bar{x}_1 >0$ and $\bar{v}>0$; the latter inequality assures that the average volume
actually contains $V_0$ in any possible equilibrium configuration.
Moreover, Eq.~(\ref{x_0_open}) has two positive solutions if $0< z<z_0$ with $z_0=1/(2e)\approx0.1839$, while no real solution exists for $z>z_0$. 
Notice that the smallest of the roots of Eq.~(\ref{x_0_open}) is that corresponding to $0<\bar{x}_0<1/2$.
This implies that the fugacity cannot be arbitrarily large and that
\begin{equation}
0< p<z<z_0. 
\end{equation}
Therefore, we conclude that equilibrium configurations in the unconstrained ensemble can indeed be realized.
In addition, we stress that there are no different states minimizing the replica energy for the same control parameters in the equilibrium region,
hence the model has no phase transitions in the unconstrained ensemble. Furthermore, it is clear that equations (\ref{x_1_open}) and (\ref{v})
are not well defined when $b=0$. As a consequence, the Thirring model ($b=0$) does not attain equilibrium states in this ensemble; in this
case, $T$, $P$, and $\mu$ cannot be taken as independent control parameters, just as it happens for macroscopic systems with short-range interactions.
We will come back to this point in Sec.~\ref{thermodynamic_relations}. Interestingly, we note that the condition $b<0$ means that the interactions within the outer parts of the system must by repulsive
to guarantee equilibrium configurations. This prevents the system from collapsing under completely open conditions in the appropriate range of control parameters.

\begin{figure}
\includegraphics[scale=1]{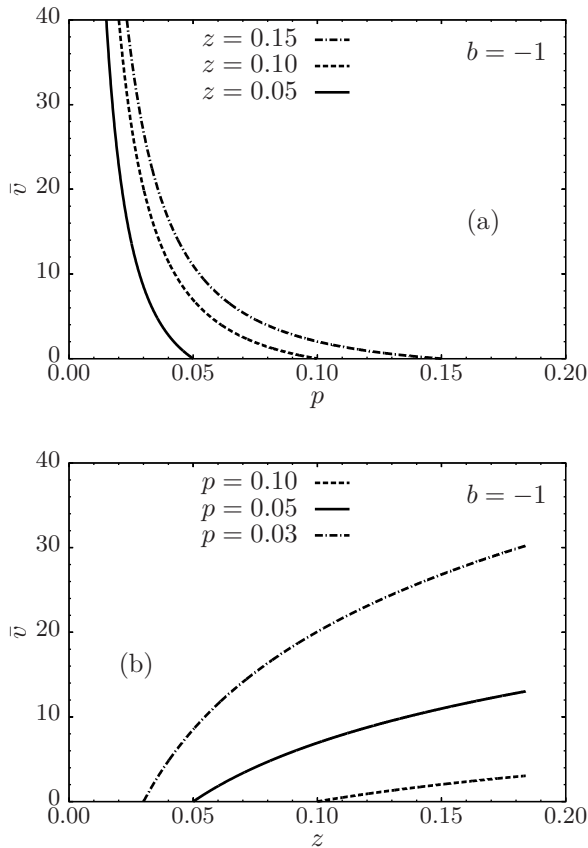}
\caption{Reduced volume $\bar{v}$ for the modified Thirring model with $b=-1$ in the unconstrained ensemble:
in (a), as a function of the reduced pressure $p$ with constant fugacity $z$, and, in (b), as a function of the fugacity with constant reduced pressure.}
\label{graph_p_v_z}
\end{figure}

In order to get further insight in the nature of the system, let us introduce the reduced number of particles $\bar{x}$ and reduced density
$\bar{y}$ defined as
\begin{eqnarray}
\bar{x}(p,z)&=&\frac{\nu \bar{N}}{T}=\bar{x}_0+\bar{x}_1,\label{x_open}\\
\bar{y}(p,z)&=&\frac{\nu V_0}{T}\frac{\bar{N}}{\bar{V}}=\frac{\bar{x}_0+\bar{x}_1}{\bar{v}+1}\label{y_open}.
\end{eqnarray}
The behavior of these thermodynamic functions in the unconstrained ensemble can be quantitatively described for the model we are studying. 
Thus, the reduced number of particles $\bar{x}(p,z)$ is shown in Fig.~\ref{graph_p_x_z}(a) as a function of the reduced pressure $p$ with constant
fugacity $z$, while in Fig.~\ref{graph_p_x_z}(b) is plotted as a function of the fugacity with constant reduced pressure.
In Fig.~{\ref{graph_p_y_z}}(a), the reduced density $\bar{y}(p,z)$ is represented as a function of the reduced pressure $p$ with constant
fugacity $z$, and, in Fig.~{\ref{graph_p_y_z}}(b), as a function of the fugacity with constant reduced pressure. Furthermore, in
Fig.~\ref{graph_p_v_z}(a), we plot the reduced volume $\bar{v}(p,z)$, as given by Eq.~(\ref{v}), as a function of the reduced pressure $p$ for
fixed values of the fugacity $z$. In Fig.~\ref{graph_p_v_z}(b), we show $\bar{v}$ as a function of $z$ holding the pressure constant.
In all these plots we have set the parameter $b$ of the model to $b=-1$.

\begin{figure}
\includegraphics[scale=1]{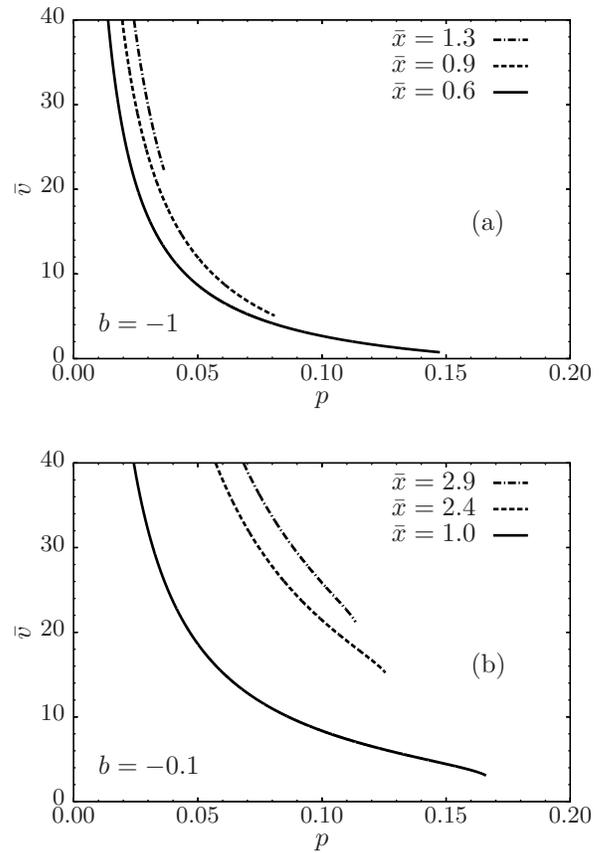}
\caption{(a) Reduced volume $\bar{v}$ as a function of the reduced pressure $p$ in the unconstrained ensemble, where $\bar{x}$ is held constant and $b=-1$.
(b) Reduced volume as a function of the reduced pressure at constant $\bar{x}$, but with $b=-0.1$.}
\label{graph_p_v_x}
\end{figure}

Let us also consider the dimensionless response functions $q_V$ and $q_N$ defined according to
\begin{eqnarray}
\frac{1}{q_V}&\equiv&-\frac{T^2}{\nu V_0^2}\left(\frac{\partial \bar{V}}{\partial P}\right)_{T,\bar{N}}
=-\left(\frac{\partial \bar{v}}{\partial p}\right)_{\bar{x}}, \\
\frac{1}{q_N}&\equiv&\nu\left(\frac{\partial \bar{N}}{\partial\mu}\right)_{T,\bar{V}}
=z(\bar{v}+1)\left(\frac{\partial \bar{y}}{\partial z}\right)_{\bar{v}}.
\end{eqnarray} 
These response functions will be useful to compare the unconstrained ensemble with the canonical ensemble (see below), and we note that $q_V$ is
related to the isothermal compressibility $\kappa_T$, according to
\begin{equation}
\frac{1}{q_V}=\frac{VT^2}{\nu V_0^2}\kappa_T.
\end{equation}
In fact, here we are only interested in the signs of $q_V$ and $q_N$. Since $V$ and $P$, and $N$ and $\mu$ are conjugate variables, respectively,
and both the number of particles and the volume fluctuate here, the response functions $q_V$ and $q_N$ must be positive in this ensemble. Moreover,
the signs of $q_V$ and $q_N$ can be seen by plotting the curves $\bar{v}$ vs. $p$ with constant $\bar{x}$, and $\bar{y}$ vs. $z$ with constant
$\bar{v}$, respectively.
Since $\bar{x}$ is not a control parameter, the curve $\bar{v}$ vs. $p$ with constant $\bar{x}$ represents the evolution of $\bar{v}$ as a function of $p$ through a series of equilibrium states where the actual control parameters, $p$ and $z$ in this case, are chosen in such a way that $\bar{x}$ takes the same value in all these states.
This can be achieved if the reduced pressure is parametrized as
\begin{equation}
p_{\bar{x}}(z)=z\exp\left\{2 b[\bar{x}-\bar{x}_0(z)]\right\}
\label{p_v_fixed}
\end{equation}
for the given $\bar{x}$, where we have used Eqs.~(\ref{x_1_open}) and (\ref{x_open}). Moreover, since the reduced pressure is always lower than $z$ in equilibrium configurations, Eq.~(\ref{p_v_fixed}) must be restricted only to values of $z$ satisfying the condition $b[\bar{x}-\bar{x}_0(z)]<0$. Hence, the curve $\bar{v}$ vs. $p$ with constant $\bar{x}$ is obtained by parametrically plotting $p_{\bar{x}}(z)$ and $\bar{v}(p_{\bar{x}}(z),z)$ using $z$ as a parameter.
The curve $\bar{y}$ vs. $z$ with constant $\bar{v}$ can be obtained in analogous manner by choosing $p$ and $z$ in such a way that $\bar{v}$ takes always the same value. In this case the reduced pressure is given by $p=p_{\bar{v}}(z)$,
where, from Eq.~(\ref{v}), $p_{\bar{v}}(z)$ is the (numerical) solution of the equation
\begin{equation}
\bar{v}=\frac{1}{2b p_{\bar{v}}}\ln\left(\frac{p_{\bar{v}}}{z}\right)
\end{equation}
with fixed values of the reduced volume $\bar{v}$.
As the reduced density $\bar{y}$ is given by Eq.~(\ref{y_open}) as a function of $p$ and $z$, the curve $\bar{y}$ vs. $z$ with constant $\bar{v}$ is obtained from Eq.~(\ref{y_open}) using $p=p_{\bar{v}}(z)$.

\begin{figure}
\includegraphics[scale=1]{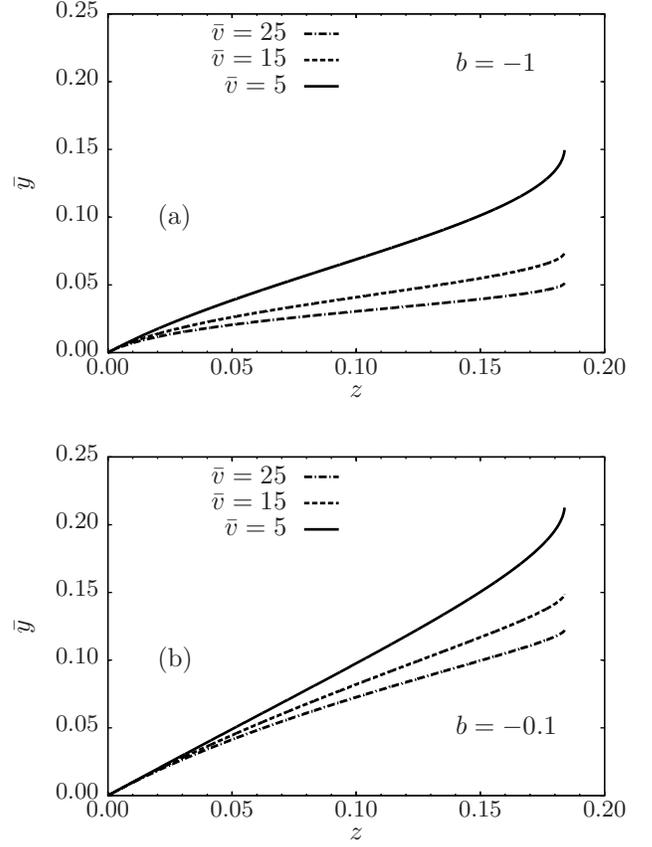}
\caption{(a) Reduced density $\bar{y}$ as a function of the fugacity $z$ in the unconstrained ensemble, where the reduced volume $\bar{v}$ is
held constant and $b=-1$. (b) Reduced density as a function of the fugacity at constant $\bar{v}$, but with $b=-0.1$.}
\label{graph_z_y_v}
\end{figure}

According to the previous discussion, in Fig.~\ref{graph_p_v_x}(a)-(b) we observe that $\bar{v}$ is a decreasing function of $p$ when $\bar{x}$ is
fixed. This means that $q_V$ is positive for these configurations, as expected. Furthermore, as can be seen in Fig.~\ref{graph_z_y_v}(a)-(b),
$\bar{y}$ increases for increasing $z$ with fixed $\bar{v}$, which indicates that the response function $q_N$ is also positive.
Below we will show that the response functions in this model can be negative in the canonical ensemble, where $V$ and $N$ are fixed control parameters.

\subsection{\label{grandcanonical_model}The grand canonical ensemble}

We now constrain the system by fixing the volume, so that the control parameters in this case are $T$, $V$, and $\mu$, which corresponds to the
grand canonical ensemble. Using Eq.~(\ref{partition_fucntion}), the grand canonical partition function $\Xi=\sum_N e^{\beta\mu N} Z$ can be written as
\begin{equation}
\Xi=\sum_N\int\frac{\dif^{3N}\vect{q}}{N!}\lambda_T^{-3N} e^{-\beta\hat{W}(N_0,N_1)} e^{\beta\mu N}.
\label{grand_canonical_partition_function} 
\end{equation}
Thus, as before, replacing the integrals over positions by a sum over all possible values of the number of particles in the two regions, according
to (\ref{Thirring_method}), one gets
\begin{equation}
\Xi=\sum_{N_0,N_1}e^{-\beta\hat{\Omega}(N_0,N_1)},
\end{equation}
where, using Striling's approximation in the large $N$ limit,
\begin{eqnarray}
\hat{\Omega}(N_0,N_1)&=&\hat{W}(N_0,N_1) -T\sum_k N_k\nonumber\\*
&&+T\sum_k N_k\left[\ln\left( N_k\frac{\lambda_T^3}{V_k}\right)-\frac{\mu}{T}\right]. 
\label{grand_potential}
\end{eqnarray}
Using the saddle-point approximation, the grand potential is given by $\Omega=\inf_{\{N_0,N_1\}}\hat{\Omega}(N_0,N_1)$. The minimization with respect
to $N_0$ and $N_1$ leads to
\begin{eqnarray}
\mu&=&-2\nu \bar{N}_0+T\ln\left(\frac{\bar{N}_0}{V_0}\lambda_T^3\right),\label{mu_1_grand}\\
\mu&=&-2b\nu \bar{N}_1+T\ln\left(\frac{\bar{N}_1}{V-V_0}\lambda_T^3\right),\label{mu_2_grand}
\end{eqnarray}
where now $\bar{N}_0$ and $\bar{N}_1$, being functions of $T$, $V$, and $\mu$, are the number of particles in each region that minimize the
grand potential. The total mean number of particles is then given by $\bar{N}=\bar{N}_0+\bar{N}_1$. In addition, in the grand canonical ensemble,
the pressure is given by
\begin{equation}
P=-\left(\frac{\partial\Omega}{\partial V}\right)_{T,\mu}= -\left(\frac{\partial\hat{\Omega}}{\partial V}\right)_{T,\mu},
\end{equation}
where the expression containing $\hat{\Omega}$ must be evaluated at $N_0=\bar{N}_0$ and $N_1=\bar{N}_1$. Thus, for the modified Thirring model,
one obtains
\begin{equation}
P=\frac{T\bar{N}_1}{V-V_0}.
\label{pressure_grand}
\end{equation}

In addition, the replica energy in the grand canonical ensemble is given by~\cite{Hill_1963}
\begin{equation}
\mathscr{E}=\Omega+PV=\Omega-V\left(\frac{\partial\Omega}{\partial V}\right)_{T,\mu}.
\label{replica_energy_grand}
\end{equation}
This means that $\mathscr{E}\neq0$ if $\Omega(T,V,\mu)$ is not a linear homogeneous function of $V$, and, therefore $\mathscr{E}$ vanishes only
when the system is additive. From Eq.~(\ref{replica_energy_grand}), using Eq.~(\ref{grand_potential}) evaluated at $N_0=\bar{N}_0$ and
$N_1=\bar{N}_1$, one obtains
\begin{equation}
\mathscr{E}= -W+P^{(e)}V,
\end{equation}
with the excess pressure $P^{(e)}=P-\bar{N}T/V$. We stress that, in this case, the replica energy is a function of $T$, $V$, and $\mu$.

In order to study the equilibrium states of the system, we introduce the reduced grand potential $\varphi_G=\nu\Omega/T^2$ and the associated function
$\hat{\varphi}_G=\nu\hat{\Omega}/T^2$, which are related through
\begin{equation}
\varphi_G(\bar{x}_0,\bar{x}_1)=\inf_{\{x_0,x_1\}} \hat{\varphi}_G(x_0,x_1).
\end{equation}
Here $x_0=\nu N_0/T$ and $x_1=\nu N_1/T$, so that $\bar{x}_0=\nu \bar{N}_0/T$ and $\bar{x}_1=\nu \bar{N}_1/T$ are the corresponding quantities
that minimize $\hat{\varphi}_G$, defining thus the equilibrium configurations in the grand canonical ensemble. From Eq.~(\ref{grand_potential}),
we obtain
\begin{eqnarray}
\hat{\varphi}_G(x_0,x_1)&=&x_0\left[\ln \left(\frac{x_0}{z}\right)-1\right]-x_0^2\nonumber\\*
&&+ x_1\left[\ln\left( \frac{x_1}{zv}\right)-1\right]-bx_1^2, 
\label{reduced_grand_potential}
\end{eqnarray}
with the relative fugacity $z=e^{(\mu-\mu_0)/T}$ and the reduced volume $v=(V-V_0)/V_0$. Here the variables $z$ and $v$ can be
taken as control parameters together with $T$. We note that, analogously to what occurred for the reduced replica energy in Sec.~\ref{completely_open_model}, the
temperature $T$ does not appear explicitly in the reduced grand potential. In dimensionless variables, Eqs.~(\ref{mu_1_grand})
and (\ref{mu_2_grand}) can be rewritten as
\begin{eqnarray}
\bar{x}_0&=&ze^{2 \bar{x}_0},\label{x0_grand}\\
\bar{x}_1&=&zve^{2b \bar{x}_1}.\label{x1_grand}
\end{eqnarray}
Furthermore, the Hessian matrix $H_G$ associated to $\hat{\varphi}_G$ at the stationary point $(\bar{x}_0,\bar{x}_1)$ takes the form
\begin{equation}
H_G=
\begin{pmatrix}
 1/\bar{x}_0-2 & 0\\
 0 & 1/\bar{x}_1-2b
\end{pmatrix},
\end{equation}
and, therefore, one finds that $\hat{\varphi}_G$ can be minimized if
\begin{equation}
\bar{x}_0<1/2,\qquad 1/\bar{x}_1>2b.
\label{conditions_grand}
\end{equation}
We thus observe that, as before, Eq.~(\ref{x0_grand}) has two solutions if $0< z<z_0$ with $z_0=1/(2e)$, and that the smallest of the roots
of Eq.~(\ref{x0_grand}) is that corresponding to $0<\bar{x}_0<1/2$. Besides that, on the one hand, Eq.~(\ref{x1_grand}) has always one solution
if $b\leq0$. On the other hand, when $b>0$, Eq.~(\ref{x1_grand}) has two solutions if $0< z<z_1(v)$, where 
\begin{equation}
z_1(v)=\frac{1}{2ebv},
\end{equation}
in such a way that the smallest of these solutions satisfies the condition (\ref{conditions_grand}). At $z=z_1(v)$, the only solution is
given by $\bar{x}_1=1/(2b)$ and hence, it corresponds to an unstable state. Therefore, Eqs.~(\ref{x0_grand}) and (\ref{x1_grand}) can be
solved simultaneously to give equilibrium configurations if 
\begin{equation}
0< z<z_G(v),
\label{z_grand}
\end{equation} 
where
\begin{equation}
z_G(v)=
\begin{cases}
z_0&\text{if }b\leq0 \\
\min[z_0,z_1(v)]&\text{if }b>0
\end{cases}.
\end{equation}

We note that for given values of the control parameters, the saddle-point equations define only one state, and, hence, there are no phase
transitions in the grand canonical ensemble. We also remark that the Thirring model ($b=0$) attains equilibrium states in the grand canonical ensemble
if $0< z<z_0$.

\begin{figure}
\includegraphics[scale=1]{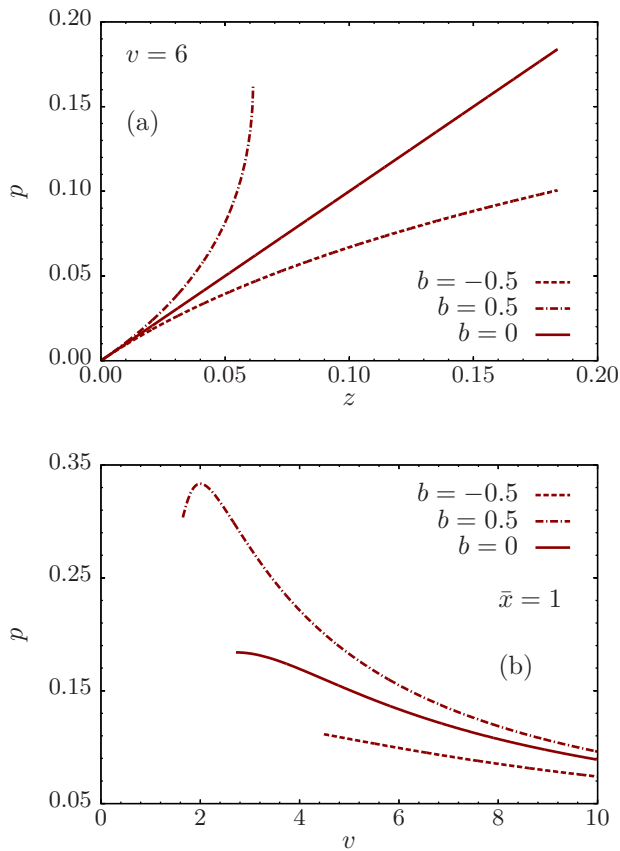}
\caption{Modified Thirring model in the grand canonical ensemble. (a) Reduced pressure as a function of the fugacity for constant reduced
volume $v$. For $b<0$, the reduced pressure is always smaller than the fugacity, while, for $b>0$, it is shown that $p>z$. In the case $b=0$,
the condition $p=z$ is always satisfied. (b) Pressure as a function of the volume with constant reduced number of particles $\bar{x}$. In the
case $b=0.5$, the portion of the curve with positive slope correspond to states of negative isothermal compressibility.}
\label{graph_grand}
\end{figure}

In the grand canonical ensemble, we now introduce the reduced number of particles, density, and pressure given by
\begin{eqnarray}
\bar{x}(v,z)&=&\frac{\nu \bar{N}}{T}=\bar{x}_0+\bar{x}_1,\label{x_grand}\\
\bar{y}(v,z)&=&\frac{\nu V_0}{T}\frac{\bar{N}}{V}=\frac{\bar{x}_0+\bar{x}_1}{v+1}\label{y_grand},\\
p(v,z)&=&\frac{\nu V_0P}{T^2}= \frac{\bar{x}_1}{v}\label{p_grand},
\end{eqnarray}
respectively. In addition, the response functions are given by
\begin{eqnarray}
q_V&=&-\frac{\nu V_0^2}{T^2}\left(\frac{\partial P}{\partial V}\right)_{T,\bar{N}}
=-\left(\frac{\partial p}{\partial v}\right)_{\bar{x}}, \label{q_V_grand}\\
\frac{1}{q_N}&=&\nu\left(\frac{\partial\bar{N}}{\partial\mu}\right)_{T,V}
=z(v+1)\left(\frac{\partial \bar{y}}{\partial z}\right)_v\label{q_N_grand}.
\end{eqnarray}
Hence, in view of Eq.~(\ref{q_V_grand}), to put in evidence the sign of $q_V$, one has to plot $p$ as function of $v$ by holding $\bar{x}$ constant.
Since $\bar{x}$ is not a control parameter in the grand canonical ensemble, the curve $p$ vs. $v$ with constant $\bar{x}$ represents the evolution of $p$ as a function of $v$ through a series of equilibrium states characterized by the same $\bar{x}$. By combining Eqs.~(\ref{x0_grand}) and (\ref{x1_grand}), $z$ can in fact be chosen in such a way that $\bar{x}$ remains constant when $v$ is varied. Using these values of $z$ in Eq.~(\ref{p_grand}), we obtain the curve $p$ vs. $v$ with constant $\bar{x}$.
Moreover, the sign of $q_N$ can be directly seen by plotting $z$ vs. $\bar{y}$ at constant $v$, since $\bar{y}=\bar{y}(v,z)$ as given by Eq.~(\ref{y_grand}).

For $b<0$, the pressure-volume relation is invertible in the grand canonical ensemble. That is, from Eqs.~(\ref{x1_grand})
and (\ref{p_grand}), we can write
\begin{equation}
v=\frac{1}{2bp(v,z)}\ln\left[\frac{p(v,z)}{z}\right], 
\end{equation}
which for constant $z$ defines the same relation as in the unconstrained ensemble. Moreover, the equilibrium conditions are the same in the
two ensembles for $b<0$. Therefore, in the modified Thirring model, the grand canonical and unconstrained ensembles are equivalent when $b<0$.

Since also the class of models with $b\geq0$ attains equilibria in the grand canonical ensemble, the phenomenology in this case is richer than in the
unconstrained case. For instance, equilibrium configurations with $p\geq z$ or with negative isothermal compressibility cannot be observed under completely open conditions, while these configurations can be realized with fixed volume. In Fig~\ref{graph_grand}(a), we show $p$ as
a function of $z$ with fixed $v$ for different values of the parameter $b$. It can be seen that $p>z$ when $b>0$, while $p<z$ when $b<0$ as it
happens in the unconstrained ensemble. In addition, in this plot we observe that $p=z$ for $b=0$, a general feature of the Thirring model. All the curves in Fig~\ref{graph_grand}(a) finish at $z=z_G(v)$, since beyond this critical fugacity the stability is lost in the grand canonical ensemble.
In Fig~\ref{graph_grand}(b), $p$ is plotted as a function of $v$ by holding $\bar{x}$ constant, where also different values of $b$ are chosen. Since holding $\bar{x}$ constant determines the value of the fugacity when $v$ is varied, the curves start at the minimum value of $v$ for which the condition $0< z<z_G(v)$ is satisfied, ensuring thus the equilibrium of the configurations.
As a remarkable fact, a region where $q_V<0$ is observed for $b=0.5$, which corresponds to the points of the curve with positive slope.
Configurations in such a region have negative isothermal compressibility. 

\subsection{\label{canonical_model}The canonical ensemble}

In order to understand better the behavior of the system in the unconstrained ensemble, it is instructive to compare its equilibrium states with the corresponding ones in the canonical ensemble. Hence, we consider now that the control parameters are $T$, $V$, and $N$.
Using (\ref{Thirring_method}) and the integral representation
\begin{equation}
\delta_{N,N_0+N_1}= \int_{\alpha-\ii\pi}^{\alpha+\ii\pi}\frac{\dif \zeta}{2\pi\ii}\ e^{\zeta \left(N-N_0-N_1\right)} 
\end{equation}
with $\text{Re}[\zeta]=\alpha=-\mu/T$, the canonical partition function (\ref{partition_fucntion}) can be written as
\begin{equation}
Z=\int_{\alpha-\ii\pi}^{\alpha+\ii\pi}\frac{\dif \zeta}{2\pi\ii}\sum_{N_0,N_1}\ \ee^{- \hat{F}(\zeta,N_0,N_1)/T},
\end{equation}
where, for large $N$,
\begin{eqnarray}
\hat{F}(\zeta,N_0,N_1)&=& -\zeta T \left(N-\sum_kN_k\right)+ \hat{W}(N_0,N_1)\nonumber\\*
&&+T\sum_kN_k \left[\ln\left(\frac{N_k}{V_k}\lambda_T^3\right)-1\right].
\label{hat_A} 
\end{eqnarray}
The canonical Helmholtz free energy is thus given by $F=\inf_{\{\alpha,N_0,N_1\}}\hat{F}(\alpha,N_0,N_1)$, where $\zeta$ is evaluated at its real part $\alpha$. Moreover, minimization with respect
to $\alpha$ enforces that $\bar{N}_1=N-\bar{N}_0$, while minimization with respect to $N_0$ and $N_1$ leads to
\begin{eqnarray}
\mu&=&-2\nu \bar{N}_0+T\ln\left(\frac{\bar{N}_0}{V_0}\lambda_T^3\right),\label{alpha_1_canonical}\\
\mu&=&-2\nu b(N-\bar{N}_0)+T\ln\left(\frac{N-\bar{N}_0}{V-V_0}\lambda_T^3\right),\label{alpha_2_canonical}
\end{eqnarray}
where we have used that $\alpha=-\mu/T$, and where the bars denote that the corresponding quantity minimizes the canonical free energy.
In view of
\begin{equation}
\mu=\left(\frac{\partial F}{\partial N}\right)_{T,V} =\left(\frac{\partial \hat{F}}{\partial N}\right)_{T,V},
\end{equation}
with the expression containing $\hat{F}$ evaluated at $N_0=\bar{N}_0$ and $N_1=\bar{N}_1$,
we note that $\mu$ is indeed the chemical potential of the system, that in the canonical ensemble is no more a control parameter. Equating the
right hand sides of Eqs. (\ref{alpha_1_canonical}) and (\ref{alpha_2_canonical}) we get an equation for $\bar{N}_0$ as a function of $T$, $V$ and $N$.
Furthermore, the pressure is given by
\begin{equation}
P=-\left(\frac{\partial F}{\partial V}\right)_{T,N} =-\left(\frac{\partial \hat{F}}{\partial V}\right)_{T,N},
\end{equation}
so that
\begin{equation}
P=\frac{T\bar{N}_1}{V-V_0}. 
\end{equation}

\begin{figure}
\includegraphics[scale=1]{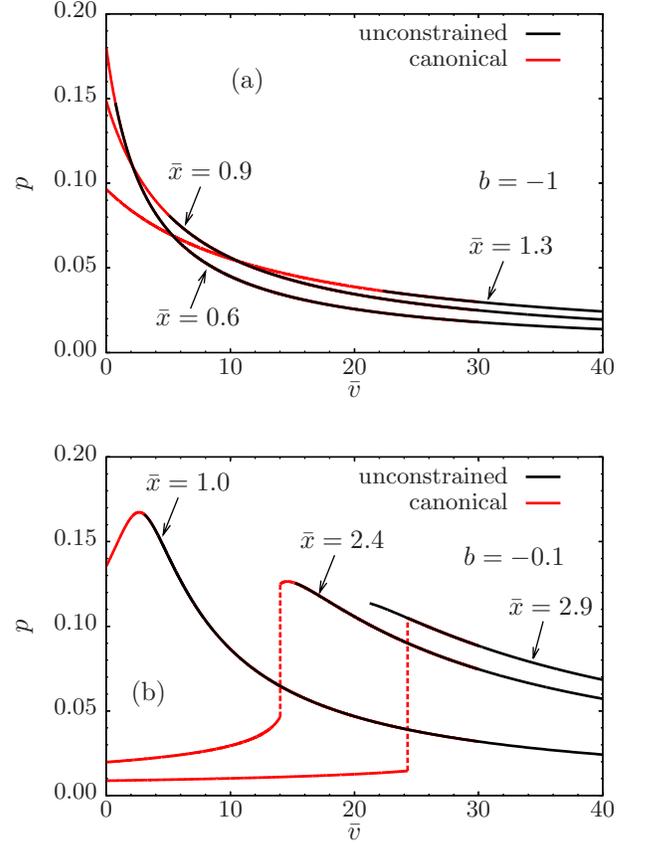}
\caption{Comparison between the unconstrained and the canonical ensembles in the modified Thirring model. (a) The black lines indicate the
reduced pressure $p$ as a function of the reduced volume $\bar{v}$ in the unconstrained ensemble, where the reduced number of particles
$\bar{x}$ is held constant for $b=-1$. The red lines show the analogous curves in the canonical ensemble, where $v$ and $x$ are fixed to
the same values of $\bar{v}$ and $\bar{x}$, respectively. (b) The plot shows the same as in (a), but with $b=-0.1$ and different values of
$\bar{x}$. For $b=-0.1$ the model shows a first-order phase transition in the canonical ensemble, as can be appreciated in the plot from the jumps in pressure (red dotted lines). In
both (a) and (b), the curves in the canonical ensemble continue to the right superposed upon the curves in the unconstrained ensemble.
We remark that in (b), the three curves in the canonical ensemble present a portion with positive slope, indicating negative compressibility. In particular, the slopes of the curves with $\bar{x}=2.4$ and $\bar{x}=2.9$ are positive before the jump.}
\label{graph_v_p_x}
\end{figure}

\begin{figure}
\includegraphics[scale=1]{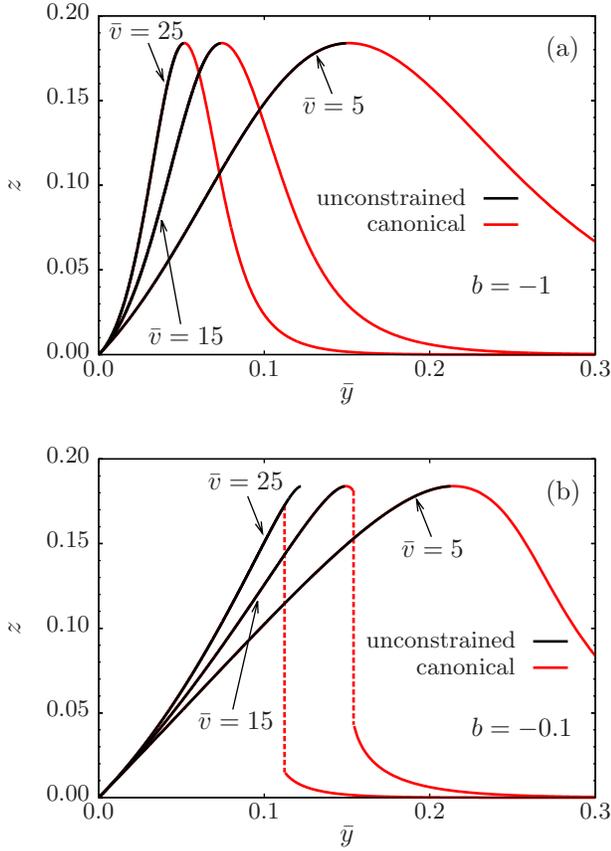}
\caption{Comparison between the unconstrained and the canonical ensembles in the modified Thirring model. (a) The black lines represent
the fugacity $z$ as a function of the reduced density $\bar{y}$ in the unconstrained ensemble. Here the reduced volume $\bar{v}$ is held
constant and we set $b=-1$. The red lines show the analogous curves in the canonical ensemble, where $y$ and $v$ are fixed to the same
values of $\bar{y}$ and $\bar{v}$, respectively. In (b), the plot shows the same as in (a), but with $b=-0.1$. The jumps correspond to
first-order phase transitions in the canonical ensemble (red dotted lines). In both (a) and (b), the curves in the canonical ensemble continue to the left
superposed upon the curves in the unconstrained ensemble.}
\label{graph_y_z_v}
\end{figure}

We now turn our attention to the replica energy in the canonical ensemble. It is given by~\cite{Hill_1963}
\begin{eqnarray}
\mathscr{E}&=&F+PV-\mu N\nonumber\\
&=&F-V\left(\frac{\partial F}{\partial V}\right)_{T,N}-N\left(\frac{\partial F}{\partial N}\right)_{T,V},
\label{replica_energy_canonical}
\end{eqnarray}
so that now $\mathscr{E}\neq0$ if $F(T,V,N)$ is not a linear homogeneous function of $V$ and $N$, and it vanishes only when the system is
additive. From Eqs.~(\ref{replica_energy_canonical}) and (\ref{hat_A}), we again obtain
\begin{equation}
\mathscr{E}= -W+P^{(e)}V,
\end{equation}
where $P^{(e)}=P-NT/V$.

Furthermore, taking $x= \nu N/T$ and $v=(V-V_0)/V_0$ as control parameters, and, as before, introducing $\bar{x}_0=\nu\bar{N}_0/T$ in the
canonical ensemble, equations (\ref{alpha_1_canonical}) and (\ref{alpha_2_canonical}) can be combined into
\begin{equation}
2bx-2(1+b)\bar{x}_0+\ln\left(\frac{\bar{x}_0}{x-\bar{x}_0}\right)+\ln v=0,
\label{canonical_eq}
\end{equation}
Notice that $\bar{x}_0/x=\bar{N}_0/N$
represents the fraction of particles inside the volume $V_0$, so that $0\leq\bar{x}_0/x\leq1$. Equation (\ref{canonical_eq}) defines
$\bar{x}_0=\bar{x}_0(x,v)$ in this ensemble, and, depending on the parameters, it can have two solutions. Again, in reduced variables the temperature does not appear. The solution determining the
equilibrium states corresponds to $x_0=\bar{x}_0$ that minimizes the canonical free energy, or, equivalently, the reduced free
energy $\varphi_C= \nu F/T^2$. From Eq.~(\ref{hat_A}) and according to the variational problem for this case, one obtains 
\begin{equation}
\varphi_C(\bar{x}_0)=\inf_{x_0}\hat{\varphi}_C(x_0) 
\label{reduced_canonical_free_energy_1}
\end{equation}
with
\begin{eqnarray}
\hat{\varphi}_C(x_0)&=&2bxx_0 -(1+b)x_0^2+x_0 \ln\left(\frac{x_0}{x-x_0}\right)\nonumber\\*
&&+x \ln\left(1-\frac{x_0}{x}\right)+x_0\ln v, 
\label{reduced_canonical_free_energy}
\end{eqnarray}
where in Eq.~(\ref{reduced_canonical_free_energy}) we have used that $\bar{N}_1=N-\bar{N}_0$, and omitted terms that do not depend on $x_0$.
It is interesting to observe that Eq.(\ref{canonical_eq}) can have more than one solution, so that the model may exhibit phase transitions.
In the Appendix we discuss how to obtain the critical point for the modified Thirring model in the canonical ensemble.

Once (\ref{canonical_eq}) is solved satisfying (\ref{reduced_canonical_free_energy_1}), the relative fugacity $z= e^{(\mu-\mu_0)/T}$ can
be computed from Eq.~(\ref{alpha_1_canonical}), since this equation can be rewritten as 
\begin{equation}
z(x,v)=\bar{x}_0e^{-2\bar{x}_0}. 
\end{equation}
Furthermore, the reduced pressure $p=\nu V_0P/T^2$ takes the form
\begin{equation}
p(x,v)= \frac{x-\bar{x}_0}{v}.
\end{equation}

The response functions, as defined in Sec.~\ref{completely_open_model}, in the canonical ensemble are given by
\begin{eqnarray}
q_V&=&\frac{\nu V_0^2}{T^2}\left(\frac{\partial^2F}{\partial V^2}\right)_{T,N}=-\left(\frac{\partial p}{\partial v}\right)_x, \\
q_N&=&\frac{1}{\nu}\left(\frac{\partial^2F}{\partial N^2}\right)_{T,V}=\frac{1}{z(v+1)}\left(\frac{\partial z}{\partial y}\right)_v,
\end{eqnarray}
where we have introduced the reduced density 
\begin{equation}
y=\frac{\nu V_0}{T}\frac{N}{V}=\frac{x}{v+1}. 
\end{equation}

As discussed in Sec.~\ref{completely_open_model}, the sign of the response functions $q_V$ and $q_N$ can be be inferred from the slopes of
the curves $p(v)$ at constant $x$ and $z(y)$ at constant $v$, respectively. Here we want to compare these curves with the corresponding
ones in the unconstrained ensemble. We do this for negative values of $b$, for which the unconstrained and the grand canonical ensembles
are equivalent. Then, the curves represent also the comparison between the canonical and grand canonical ensembles when $b<0$. Thus, in
Fig.~\ref{graph_v_p_x}(a) we show $p$ as a function of $\bar{v}$ in the unconstrained ensemble
with $b=-1$, where $\bar{x}$ is held constant. The corresponding curves in the canonical ensemble are also shown in this plot, where
$v$ and $x$ are fixed to the same values of $\bar{v}$ and $\bar{x}$, respectively. In Fig.~\ref{graph_v_p_x}(b), these curves are
represented for $b=-0.1$ and different values of $\bar{x}$, where it can be appreciated that the model presents first-order phase
transitions in the canonical ensemble, as indicated by the jumps in the pressure. In addition, in Fig.~\ref{graph_y_z_v}(a) we represent
$z$ as a function of $\bar{y}$ in the unconstrained ensemble, where $\bar{v}$ is held constant and we set $b=-1$. We also show the
analogous curves in the canonical ensemble, where $y$ and $v$ are fixed to the same values of $\bar{y}$ and $\bar{v}$, respectively.
In Fig.~\ref{graph_y_z_v}(b), these curves are shown for $b=-0.1$. The jumps in the fugacity correspond to first-order phase transitions
in the canonical ensemble. Thus, the canonical and unconstrained ensembles are nonequivalent at the macrostate level~\cite{Touchette_2004,Touchette_2015}, since the equilibirum states in the two ensembles are not in one-to-one correspondence.
In particular, we highlight that the response functions can be negative in the canonical ensemble.
Finally, we briefly note that a nonequivalence between the microcanonical and canonical ensembles is expected for the modified Thirring model, since this happens for the particular case $b=0$~\cite{Campa_2016}.

\section{\label{thermodynamic_relations}Thermodynamic relations and replica energy}
 
In Sec.~\ref{unconstrained_ensemble}, from an equation for the differential variations of the replica energy in the unconstrained ensemble,
we have obtained a set of relations in terms of partial derivatives that allows one to obtain, e.g., the entropy of the system.
Let us recall these relations since they will be the central issue of the following discussion:
\begin{equation}
\dif\mathscr{E}=-S\dif T+ V\dif P-N\dif\mu,
\label{generalized_Gibbs_Duhem_2}
\end{equation}
and
\begin{eqnarray}
\left(\frac{\partial\mathscr{E}}{\partial T}\right)_{P,\mu}&=&-S,\label{entropy_relation_4}\\ 
\left(\frac{\partial\mathscr{E}}{\partial P}\right)_{T,\mu}&=&V,\label{volume_relation_4}\\
\left(\frac{\partial\mathscr{E}}{\partial \mu}\right)_{T,P}&=&-N.\label{N_relation_4}
\end{eqnarray}
These equations, however, are thermodynamic relations valid in any ensemble; with this in mind, we do not use here a bar over the variables
to indicate whether a certain quantity fluctuates or not.

The present discussion is to emphasize that a situation may exist in which the replica energy is different from zero, but $T$, $P$, and
$\mu$ cannot be taken as a set of independent variables. This can happen in an ensemble different from the unconstrained one. When
$T$, $P$, and $\mu$ cannot be taken as independent variables in the unconstrained ensemble, the mean-field equations will not lead to a minimum
of replica energy, and, therefore, Eqs.~(\ref{generalized_Gibbs_Duhem_2})-(\ref{N_relation_4}) are meaningless.

Let us assume that the system is itself constrained such that, for instance, $\mu=\mu(T,P)$. 
In this case, Eq.~(\ref{generalized_Gibbs_Duhem_2}) becomes
\begin{equation}
\dif\mathscr{E}=-\left[S+N\left(\frac{\partial\mu}{\partial T}\right)_P\right]\dif T+ \left[V-N\left(\frac{\partial\mu}{\partial P}\right)_T\right]\dif P,
\label{generalized_Gibbs_Duhem_3}
\end{equation}
which establishes a functional relation $\mathscr{E}=\mathscr{E}(T,P)$.
Thus, one actually has
\begin{eqnarray}
\left(\frac{\partial\mathscr{E}}{\partial T}\right)_{P}&=&-S-N\left(\frac{\partial\mu}{\partial T}\right)_P,\label{entropy_relation_2}\\ 
\left(\frac{\partial\mathscr{E}}{\partial P}\right)_{T}&=&V-N\left(\frac{\partial\mu}{\partial P}\right)_T,\label{volume_relation_2}
\end{eqnarray}
which are the relations satisfied in this case instead of Eqs.~(\ref{entropy_relation_4}) and (\ref{volume_relation_4}). This is equivalent
to directly consider a constraint $\mathscr{E}=\mathscr{E}(T,P)$ in Eq.~(\ref{generalized_Gibbs_Duhem_2}), which yields
\begin{equation}
N\dif\mu=-\left[S+\left(\frac{\partial\mathscr{E}}{\partial T}\right)_P\right]\dif T+ \left[V-\left(\frac{\partial\mathscr{E}}{\partial P}\right)_T\right]\dif P.
\end{equation}
The above expression shows that the usual Gibbs-Duhem equation is not valid when the replica energy is different from zero
{\it even if} $T$, $P$,
and $\mu$ are not a set of independent variables. Moreover, analogous equations can be obtained if one considers any other functional relation
constraining $T$, $P$, and $\mu$.

To go further, consider an arbitrary $d$-dimensional system with long-range interactions whose number density at a point $\vect{x}\in\mathbb{R}^d$
is given by 
\begin{equation}
n(\vect{x})=\frac{1}{\lambda_T^d}\exp\left[\frac{\mu-\Phi(\vect{x})}{T}\right], 
\label{number_density}
\end{equation}
where $\Phi(\vect{x})$ is the mean-field potential. 
The chemical potential and the entropy of the system satisfy~\cite{Latella_2015}
\begin{eqnarray}
\mu N&=&T\int n(\vect{x})\ln\left[n(\vect{x})\lambda_T^d\right]\dif^d\vect{x}+2W,\label{mu_mean-field}\\
-S&=&\int n(\vect{x})\ln\left[n(\vect{x})\lambda_T^d\right]\dif^d\vect{x}-\frac{2+d}{2}N\label{entropy_mean-field},
\end{eqnarray}
where $W=\frac{1}{2}\int n(\vect{x})\Phi(\vect{x})\dif^d\vect{x}$ is the potential energy.
Moreover, the local pressure is given by $p(x)=n(x)T$ when short-range interactions are completely ignored, and the pressure $P$ is $p(x)$ evaluated at the boundary of the system~\cite{Latella_2013}.
Consider also that the mean-field potential is not completely arbitrary but it always vanishes at the boundary of the system, regardless of the
thermodynamic state of the system. This will enforce the condition
\begin{equation}
\mu= T\ln\left(\frac{P\lambda_T^d}{T}\right). 
\label{particular_chemical_potential}
\end{equation}
Thus, using Eq.~(\ref{particular_chemical_potential}) in Eqs.~(\ref{entropy_relation_2}) and (\ref{volume_relation_2}), and then using
Eqs.~(\ref{mu_mean-field}) and (\ref{entropy_mean-field}) leads to
\begin{eqnarray}
\left(\frac{\partial\mathscr{E}}{\partial T}\right)_{P}&=&-\frac{2W}{T},\label{entropy_relation_3}\\ 
\left(\frac{\partial\mathscr{E}}{\partial P}\right)_{T}&=&\frac{P^{(e)}V}{P},\label{volume_relation_3}
\end{eqnarray}
where $P^{(e)}=P-NT/V$. Hence, with the particular constraint (\ref{particular_chemical_potential}), Eqs.~(\ref{entropy_relation_3}) and
(\ref{volume_relation_3}) hold in place of Eqs.~(\ref{entropy_relation_4}) and (\ref{volume_relation_4}).

The interesting fact here is that the modified Thirring model can be used to test these general considerations. In
Sec.~\ref{completely_open_model} we found that in the case $b=0$ the model never attains equilibrium configurations in the unconstrained
ensemble. As we shall see below, this is due to the fact that, for the case $b=0$, the chemical potential satisfies
Eq.~(\ref{particular_chemical_potential}) (with $d=3$), and thus $T$, $P$, and $\mu$ cannot be taken as independent variables. But first
we will check that for $b\neq 0$ the entropy can be computed from Eq.~(\ref{entropy_relation_4}), showing that $T$, $P$, and $\mu$ can
actually be taken as independent variables in this case. This, of course, is in agreement with the statistical mechanics description of
the system obtained in Sec.~\ref{completely_open_model}.

To check the validity of Eq.~(\ref{entropy_relation_4}), we need to write the replica energy as a function of $T$, $P$, and $\mu$.
We note that $\Phi(\vect{x})=-2\nu\left[N_0\theta_{V_0}(\vect{x})+bN_1\theta_{V_1}(\vect{x})\right]$ and $N_1/V_1=P/T$ for the
modified Thirring model. Thus, evaluating the number density (\ref{number_density}) at any point $\vect{x}$ in $V_1$ and rearranging
terms gives us
\begin{equation}
\mu=T\ln\left(\frac{P\lambda_T^3}{T}\right)-2\nu bN_1,
\label{N_1}
\end{equation}
which defines $N_1=N_1(T,P,\mu)$ if $b\neq0$. Analogously, evaluating (\ref{number_density}) at any point $\vect{x}$ in $V_0$ yields
\begin{equation}
\mu=T\ln\left(\frac{N_0\lambda_T^3}{V_0}\right)-2\nu N_0, 
\label{N_0}
\end{equation}
which defines implicitly $N_0=N_0(T,\mu)$. Therefore, when $b\neq0$, we have
\begin{eqnarray}
\left(\frac{\partial N_0}{\partial T}\right)_{P,\mu}&=&\frac{2\nu N_0^2+\mu N_0-\frac{3}{2}N_0T}{\left(2\nu N_0-T\right)T},\\
\left(\frac{\partial N_1}{\partial T}\right)_{P,\mu}&=&\frac{N_1}{T}+\frac{1}{2b\nu}\left(\frac{\mu}{T}-\frac{5}{2}\right).
\end{eqnarray}

Since
\begin{eqnarray}
\mathscr{E}&=&-W+P^{(e)}V\nonumber\\
&=&\nu\left(N_0^2+bN_1^2\right)+PV_0-N_0T,
\end{eqnarray}
the replica energy is implicitly given as a function of $T$, $P$, and $\mu$ by means of Eqs.~(\ref{N_1}) and (\ref{N_0}).
Hence,
\begin{equation}
\left(\frac{\partial \mathscr{E}}{\partial T}\right)_{P,\mu}= \sum_k N_k\left[\ln\left(\frac{N_k\lambda_T^3}{V_k}\right)-\frac{5}{2}\right],
\label{minus_entropy}
\end{equation}
where we have rearranged terms using Eqs.~(\ref{N_1}) and (\ref{N_0}). By comparing with Eq.~(\ref{entropy_mean-field}), we thus see that
the rhs of Eq.~(\ref{minus_entropy}) is indeed $-S$. This confirms that when $b\neq0$, the system has enough thermodynamic degrees of
freedom to take $T$, $P$, and $\mu$ as a set of independent variables. To see whether the configurations are stable or not, however, one
must perform an analysis of the second order variations of the appropriate free energy, as we have done in Sec.~\ref{completely_open_model},
for instance, depending on the actual physical conditions imposed by the corresponding control parameters.

In view of Eq.~(\ref{N_1}), it is now obvious that if $b=0$, the chemical potential is given by Eq.~(\ref{particular_chemical_potential})
and that now
\begin{equation}
2\nu N_0=T\ln\left(\frac{N_0T}{PV_0}\right)
\label{N_0_2}
\end{equation}
defines implicitly $N_0=N_0(T,P)$.
Hence,
\begin{equation}
\left(\frac{\partial N_0}{\partial T}\right)_{P}=\frac{N_0}{T}\left(\frac{2\nu N_0+T}{2\nu N_0-T}\right).\label{partial_N0} 
\end{equation}
Using Eq.(\ref{partial_N0}) and since in this case $\mathscr{E}=\nu N_0^2+PV_0-N_0T$, it is easy to see that
\begin{equation}
\left(\frac{\partial \mathscr{E}}{\partial T}\right)_P= \frac{2\nu N_0^2}{T}=-\frac{2W}{T},
\end{equation}
in agreement with Eq.~(\ref{entropy_relation_3}). These arguments show that, although the replica energy is different from zero, in the
Thirring model ($b=0$) one cannot take $T$, $P$, and $\mu$ as a set of independent variables.

\section{\label{conclusions}Conclusions}

We have shown that systems with long-range interactions can attain configurations of thermodynamic equilibrium in the unconstrained ensemble.
In this ensemble, the control parameters are temperature, pressure, and chemical potential and the appropriate free energy is the replica
energy. We have presented a solvable model that is stable in this ensemble, and we have compared its equilibrium states with those of the
grand canonical and canonical ensembles. From this comparison we observe that the space of parameters defining the possible stable
configurations is enlarged when the system is constrained by fixing the volume and the number of particles. These quantities fluctuate in
the unconstrained ensemble, as well as the energy of the system.
Moreover, the model we have introduced exhibits first-order phase transitions in the canonical ensemble at which the system undergoes pressure and fugacity jumps.

On the one hand, macroscopic systems with short-range interactions cannot attain equilibrium states if the control parameters are the
temperature, pressure, and chemical potential. According to the usual Gibbs-Duhem equation, these variables cannot be taken as independent.
A physical reason behind this property is that in this case temperature, pressure, and chemical potential are truly intensive properties and therefore, they cannot define the size of the system corresponding to an equilibrium state. On the other hand, however, in systems with
long-range interactions temperature, pressure, and chemical potential are not intensive properties, so that controlling, e.g.,
the temperature $T\sim N$ can actually define the size of the system in an equilibrium configuration. Thus, the typical scaling of the
thermodynamic variables in long-range interacting systems makes it possible to have equilibrium configurations
in completely open conditions. 

\begin{acknowledgments}
I.L. acknowledges financial support through an FPI scholarship (BES-2012-054782) from the Spanish Government under Project FIS2011-22603.
\end{acknowledgments}

\appendix*
\section{\label{appendix_phase_transitions}Critical point of the modified Thirring model in the canonical ensemble}

Phase transitions in the modified Thirring model in the canonical ensemble can be studied by extending the analysis done in
Ref.~\cite{Campa_2016} using Landau theory of phase transitions. In~\cite{Campa_2016}, the expansion parameter specifying the
transition is taken as the deviation of the fraction $1-N_0/N$ with respect to the value of this fraction at the transition line. Here,
equivalently, we consider the fraction of particles in $V_0$, given by $x_0/x$, and take $m=(x_0-\bar{x}_0)/x$ as the expansion parameter.
Accordingly, to obtain the critical point we expand the free energy (\ref{reduced_canonical_free_energy}) as
$\hat{\varphi}_C=\varphi_0+\varphi_1m+\varphi_2m^2+\mathcal{O}(m^3)$, and look for a solution of the system of equations
\begin{equation}
\varphi_1(\bar{x}_0,x,v)=0,\qquad \varphi_2(\bar{x}_0,x,v)=0,
\label{system_phase_transitions}
\end{equation}
with the constraint $ 0\leq\bar{x}_0/x\leq1$.
The first of the equations (\ref{system_phase_transitions}) is exactly (\ref{canonical_eq}), while the second one can be written as
\begin{equation}
\frac{ 2 (b+1) (x-\bar{x}_0)\bar{x}_0-x}{ (x-\bar{x}_0) \bar{x}_0}=0.
\label{second_coef}
\end{equation}
Thus, for $b>-1$, Eq.~(\ref{second_coef}) has two real solutions, $\bar{x}_0=x_{\pm}(x)=\left[1+b\pm\sqrt{(1+b)(1+b-2/x)}\right]^{-1}$,
when $x>2/(1+b)\equiv x_c$. The critical point is defined by the condition~\cite{Campa_2016} $x_+(x_c)=x_-(x_c)$, in such a way that
at the critical point one has $x=x_c$ and $\bar{x}_0=x_\pm(x_c)=x_c/2$. In terms of the thermodynamic variables, this means that the
critical temperature is given by $T_c=\nu N(1+b)/2$ and that the fraction $\bar{N}_0/N$ at the critical point is $1/2$. Moreover, we note
that when $b=-1$, the l.h.s. of Eq.~(\ref{second_coef}) does not vanish, and, therefore, there are no phase transitions in this case.
For $b<-1$, the fractions $x_{\pm}/x$ do not lie in the interval $[0,1]$, so that also in this case the model will not exhibit phase
transitions. We therefore observe that the model may present phase transitions in the canonical ensemble, depending of the value of $v$,
if $x$ is larger than $x_c$ and if $b>-1$.

The critical value of the reduced volume $v_c$ is obtained by replacing $\bar{x}_0=x_\pm(x_c)$ in Eq.~(\ref{canonical_eq}) with $v=v_c$,
yielding $v_c=\exp[2(1-b)/(1+b)]$. Thus, the model may exhibit phase transitions when $v>v_\text{c}$.

\end{document}